\documentclass[sigconf,natbib=true]{acmart}
\pdfoutput=1
\usepackage{booktabs}
\usepackage{amsmath}
\usepackage{multirow}
\usepackage{listings} 
\usepackage{tabularx}
\usepackage{graphicx} 
\usepackage{hyperref}
\usepackage{enumitem}
\usepackage{subcaption}
\usepackage{colortbl}

\setlist[itemize]{leftmargin=*} 
\AtBeginDocument{%
	}

\copyrightyear{2025}
\acmYear{2025}
\setcopyright{acmlicensed}\acmConference[RecSys '25]{Proceedings of the Nineteenth ACM Conference on Recommender Systems}{September 22--26, 2025}{Prague, Czech Republic}
\acmBooktitle{Proceedings of the Nineteenth ACM Conference on Recommender Systems (RecSys '25), September 22--26, 2025, Prague, Czech Republic}
\acmDOI{10.1145/3705328.3748044}
\acmISBN{979-8-4007-1364-4/2025/09}

\begin{document}
	
	\title{Enhancing Transferability and Consistency in Cross-Domain Recommendations via Supervised Disentanglement}
	
	\author{Yuhan Wang}
	\orcid{0000-0003-1526-276X}
	\email{wyh0520@whut.edu.cn}
	\affiliation{%
		\institution{Wuhan University of Technology}
		\city{Wuhan}
		\country{China}
		\postcode{430070}}
	\affiliation{%
		\institution{Engineering Research Center of Intelligent Service Technology for Digital Publishing, Ministry of Education}
		\city{Wuhan}
		\country{China}
		\postcode{430070}}
	
	\author{Qing Xie}
	\orcid{0000-0003-4530-588X}
	\email{felixxq@whut.edu.cn}
	\authornote{Corresponding authors}
	\affiliation{%
		\institution{Wuhan University of Technology}
		\city{Wuhan}
		\country{China}
		\postcode{430070}}
	\affiliation{%
		\institution{Engineering Research Center of Intelligent Service Technology for Digital Publishing, Ministry of Education}
		\city{Wuhan}
		\country{China}
		\postcode{430070}}
	
	\author{Zhifeng Bao}
	\orcid{0000-0003-2477-381X}
	\email{zhifeng.bao@rmit.edu.au}
	\affiliation{%
		\institution{School of Computing Technologies, RMIT University}
		\city{Melbourne}
		\country{Australia}
		\postcode{3000}}
	
	\author{Mengzi Tang}
	\orcid{0000-0001-5611-673X}
	\email{tangmz@whut.edu.cn}
	\affiliation{%
		\institution{Wuhan University of Technology}
		\city{Wuhan}
		\country{China}
		\postcode{430070}}
	\affiliation{%
		\institution{Engineering Research Center of Intelligent Service Technology for Digital Publishing, Ministry of Education}
		\city{Wuhan}
		\country{China}
		\postcode{430070}}
	
	\author{Lin Li}
	\orcid{0000-0001-7553-6916}
	\email{cathylilin@whut.edu.cn}
	\affiliation{%
		\institution{Wuhan University of Technology}
		\city{Wuhan}
		\country{China}
		\postcode{430070}}
	\affiliation{%
		\institution{Engineering Research Center of Intelligent Service Technology for Digital Publishing, Ministry of Education}
		\city{Wuhan}
		\country{China}
		\postcode{430070}}
	
	\author{Yongjian Liu}
	\orcid{0009-0005-8388-6736}
	\email{liuyj@whut.edu.cn}
	
	\affiliation{%
		\institution{Wuhan University of Technology}
		\city{Wuhan}
		\country{China}
		\postcode{430070}}
	\affiliation{%
		\institution{Engineering Research Center of Intelligent Service Technology for Digital Publishing, Ministry of Education}
		\city{Wuhan}
		\country{China}
		\postcode{430070}}
	
	\renewcommand{\shortauthors}{Wang et al.}
	
	\begin{abstract}
		Cross-domain recommendation (CDR) aims to alleviate the data sparsity by transferring knowledge across domains. Disentangled representation learning provides an effective solution to model complex user preferences by separating intra-domain features (domain-shared and domain-specific features), thereby enhancing robustness and interpretability. However, disentanglement-based CDR methods employing generative modeling or GNNs with contrastive objectives face two key challenges: (i) \textbf{pre-separation strategies} decouple features before extracting collaborative signals, disrupting intra-domain interactions and introducing noise; (ii) \textbf{unsupervised disentanglement objectives} lack explicit task-specific guidance, resulting in limited consistency and suboptimal alignment. To address these challenges, we propose DGCDR, a GNN-enhanced encoder-decoder framework. To handle challenge (i), \textbf{DGCDR first applies GNN to extract high-order collaborative signals}, providing enriched representations as a robust foundation for disentanglement. The encoder then dynamically disentangles features into domain-shared and -specific spaces, preserving collaborative information during the separation process. To handle challenge (ii), \textbf{the decoder introduces an anchor-based supervision} that leverages hierarchical feature relationships to enhance intra-domain consistency and cross-domain alignment. Extensive experiments on real-world datasets demonstrate that DGCDR achieves state-of-the-art performance, with improvements of up to 11.59\% across key metrics. Qualitative analyses further validate its superior disentanglement quality and transferability. Our source code and datasets are available on GitHub for further comparison.\footnote{\url{https://github.com/WangYuhan-0520/DGCDR}}
	\end{abstract}

	
	
	\maketitle
	\section{Introduction}
	Cross-domain recommendation (CDR) jointly leverages available data from related domains to enhance recommender systems that may suffer from the data sparsity issue \cite{ZangZLZY23}. A fundamental assumption in CDR is that users share some common interests across different domains, stemming from their personalities \cite{LiuLLP20, ZhuTLZXZLH22, ZhangCLYP24}. Therefore, many popular CDR methods \citep{ManSJC17, KangHLY19, ZhuGZXXZL021, SalahTL21} based on Embedding and Mapping (EMCDR) framework use a collaborative filtering-based model to extract the features within each domain, and then align the overlapping features by inter-domain mapping functions. These methods rely on the premise that user preferences remain consistent across domains, and fit the data distribution by taking a shortcut strategy that directly transfers shared features. However, such an assumption oversimplifies user behaviors by neglecting the presence of domain-specific preferences that reflect users' unique interests tailored to particular domains, leading to negative transfer problems \cite{ZhangCLYP24}.
	
	To better model user behaviors and understand the diverse nature of user preferences, it is essential to disentangle these domain-shared and domain-specific preferences.  Disentangled representation learning (DRL) \cite{WangX2024, Zhu23}, originally developed in the field of computer vision, focuses on identifying the underlying factors of observed data and decomposing them into independent vector spaces. By decomposing and reconstructing the entangled features, DRL facilitates deep learning on the intrinsic physical properties of the data and enhances robustness for downstream applications. This capability makes DRL particularly appealing for recommendation systems, where the latent factors behind user behaviors are often complex and obscure \cite{WangX2024, WangJZ0XC20}, making it difficult to distinguish between domain-shared and domain-specific preferences. By enabling the separation of such entangled factors, DRL offers a principled approach to reduce complexity, enhance learning efficiency, and improve interpretability \cite{WangX2024}. 
	
	Existing CDR methods have explored disentangled representation learning with promising results, and are largely based on two paradigms: generative models like Variational Autoencoders (VAEs), and Graph Neural Networks (GNNs) combined with contrastive learning. While VAE-based methods \cite{CaoLCYL022, CaoSCLW22, ZhangCLYP24} use variational inference to disentangle intra-domain (domain-shared and domain-specific) features, they primarily rely on distribution modeling over historical interactions and often ignore the intricate user-item relationships that reflect higher-order connectivity of collaborative signals \cite{WangJZ0XC20}. In contrast, GNN-contrastive methods \cite{ZhangZZWWY23, LiuSNJ024} leverage graph-based information propagation to capture high-order user-item relationships, and enhance intra-domain features disentanglement through self-supervised constraints. However, two key challenges remain in the existing disentangled CDR methods: 
	
	\noindent \textbf{C1. Pre-separation disrupts collaboration}. Many methods \cite{ZhangZZWWY23, LiuSNJ024, Ning0LCZT23}, while leveraging GNNs to capture high-order collaborative signals, adopt a pre-separation strategy that projects initial features into isolated spaces for domain-shared and domain-specific components before GNN propagation. This isolated strategy hinders the joint modeling of complementary information during preference propagation, as domain-specific features may encode signals beneficial for shared representations.  Furthermore, incorrect pre-separation can introduce noise into the subsequent alignment, such as misclassifying specific features for shared ones, ultimately degrading representation quality and model performance, especially in scenarios with large domain discrepancies.
	
	\noindent \textbf{C2. Lack of supervisory signals for disentangled representation learning}. As demonstrated by Locatello et al. \cite{LocatelloBLRGSB19}, disentangled representation learning in unsupervised settings is highly sensitive to randomness and hyperparameters. This underscores the necessity of task-specific guidance for feature separation \cite{WangX2024}. However, most existing CDR methods adopt unsupervised disentanglement objectives. Specifically, VAE-based methods enforce the posterior of latent factors to align with standard multivariate Gaussian priors without supervision \cite{YangLCSHW21}; while GNN-contrastive methods typically employ self-supervised objectives that penalize distant positive samples (i.e., domain-shared features) and overly close negatives (i.e., intra-domain features) in the embedding space \cite{ZhangZZWWY23}. Such a lack of explicit or implicit task-oriented supervision hinders intra-domain consistency and cross-domain transferability.
	
	Therefore, we propose DGCDR, a Disentangled Graph Cross-Domain Recommendation method, which integrates GNN-enhanced feature extraction with a disentangled encoder-decoder framework. As shown in Table \ref{tab:comparison}, DGCDR offers a comprehensive solution to address the challenges of existing methods. Specifically, we summarize the main contributions as follows:
	
	\begin{table}[]
		\centering
		\caption{Qualitative comparison across CDR categories}
		\setlength{\tabcolsep}{4pt}
		\renewcommand{\arraystretch}{1.2}
		\resizebox{\linewidth}{!}{ 
			\begin{tabular}{@{}lcccc@{}}
				\toprule
				\textbf{Capability aspect} & \textbf{NoDisen} & \textbf{VAE-based} & \textbf{GNN-contrastive} & \textbf{DGCDR(ours)} \\ 
				\midrule
				User complexity modeling & \cellcolor{orange!10}+ & \cellcolor{orange!30}+ + & \cellcolor{orange!70}+ + + & \cellcolor{orange!70}+ + + \\
				High-order signal capture & \cellcolor{white} & \cellcolor{orange!10}+ & \cellcolor{orange!90}+ + + + & \cellcolor{orange!90}+ + + + \\
				Disentanglement stability & \cellcolor{white} & \cellcolor{orange!10}+ & \cellcolor{orange!10}+  & \cellcolor{orange!70}+ + + \\
				\bottomrule
			\end{tabular}
		}
		\footnotesize{\textit{Note}: NoDisen: non-disentangled methods; Color depth indicates performance level.}
		\label{tab:comparison}
	\end{table}

	\begin{itemize}
		\item For \textbf{C1}, DGCDR begins by leveraging GNN to capture high-order collaborative information within a single domain, enriching the representation of user-item relationships. This step generates comprehensive initial features, ensuring a robust foundation for the subsequent disentanglement process. Then, through a disentangled encoder, we map the GNN-enhanced features into two distinct sets, representing domain-shared and domain-specific features, respectively. Crucially, we minimize the discrepancies using similarity measures to enable matching of shared features between domains, and separate the intra-domain features by enforcing orthogonality to facilitate disentanglement. We further incorporate interaction-level specificity to refine disentangled representations, enabling a more precise and nuanced understanding of cross-domain user preferences.
		\item For \textbf{C2}, we emphasize the necessity of supervision in disentangled representation learning for cross-domain recommendation, alleviating the instability of unsupervised methods by enhancing feature separation and cross-domain alignment. Since constructing explicit supervision for disentangled representations in CDR is non-trivial and often cost-prohibitive, to address this, we draw inspiration from a modality-disentangled approach \cite{Han0NL22} and introduce a novel contrastive-based decoder with supervisory objectives. Different from existing methods, our approach designs an anchor mechanism to guide the disentanglement process directly. Specifically, we measure the hierarchical relationships between the current domain anchor and cross-domain domain-shared, domain-specific, and GNN-enhanced features, respectively. Hierarchical relationships based on appropriate mutual information are established to facilitate the precise alignment of domain-shared features and the disentanglement of intra-domain features. With explicit supervision signals, our method is able to provide more stable and reliable cross-domain recommendations.
		\item  Extensive experiments on real-world datasets show that DGCDR outperforms optimal baselines with up to 11.59\% improvement in key metrics, and visual analyses further validate its superior disentanglement capability and cross-domain transferability.
	\end{itemize}
	
	\section{Related Work}
	Disentangled Representation Learning (DRL) aims to separate the underlying factors in data into semantically meaningful representations. This learning strategy fosters human-like generalization capability and improves the explainability, controllability, and robustness of machine learning tasks \cite{WangX2024}. Evolving from traditional techniques such as Independent Component Analysis, DRL now leverages deep-learning-based generative models such as VAE and generative adversarial networks, achieving notable performance across various applications. In computer vision, DRL facilitates the disentanglement of object attributes such as shape, color, and size, thus enhancing image manipulation and intervention \cite{Dupont18, WuLS21}. In natural language processing, DRL aids in separating semantics from syntax to enhance text generation and style transfer \cite{BaoZHLMVDC19, ChengMSMZLC20}. In addition, DRL has shown great value in areas such as multimedia data analysis \cite{XuLTLHSTGD22} and graph representation learning \cite{LiWZYLZ21}.
	
	The application of DRL to recommendation tasks has attracted significant attention due to its ability to model the inherent complexity of user behaviors. User preferences are shaped by diverse latent factors such as interests, needs, and context. DRL can effectively capture such complex preferences to improve recommendation accuracy.  In cross-domain recommendation (CDR), disentangled methods separate interaction data into domain-specific and domain-shared components, enabling knowledge transfer and reducing multicollinearity \cite{Li00024}. Existing disentangled CDR methods typically employ generative models like VAE, which use variational inference to model latent factors and reconstruct user preferences \cite{CaoLCYL022, CaoSCLW22, ZhangCLYP24}. Another line of research integrates GNN with contrastive learning to capture high-order collaborative information while applying self-supervised constraints to facilitate the disentanglement process \cite{ZhangZZWWY23, LiuSNJ024}.
	
	Despite the potential of DRL in cross-domain recommendation, challenges remain in achieving effective disentanglement in unsupervised settings, particularly in ensuring robust cross-domain alignment. Furthermore, many GNN-based methods \cite{ZhangZZWWY23, LiuSNJ024, Ning0LCZT23} adopt a pre-separation strategy that isolates domain-specific and domain-shared features before collaborative modeling, limiting representation expressiveness and increasing sensitivity to noise. In contrast, our DGCDR begins by employing GNN to model high-order collaborative signals, providing a stronger foundation for disentanglement. Furthermore, we incorporate an encoder-decoder framework with an anchor mechanism to guide the disentanglement process and improve cross-domain alignment. With this model design, we aim to reduce feature redundancy and enhance feature transferability, and address the limitations of existing methods.
	
	\section{Methodology}
	\begin{figure*}[]
		\centering
		\resizebox{\linewidth}{!}{ 
			\includegraphics[width=\linewidth]{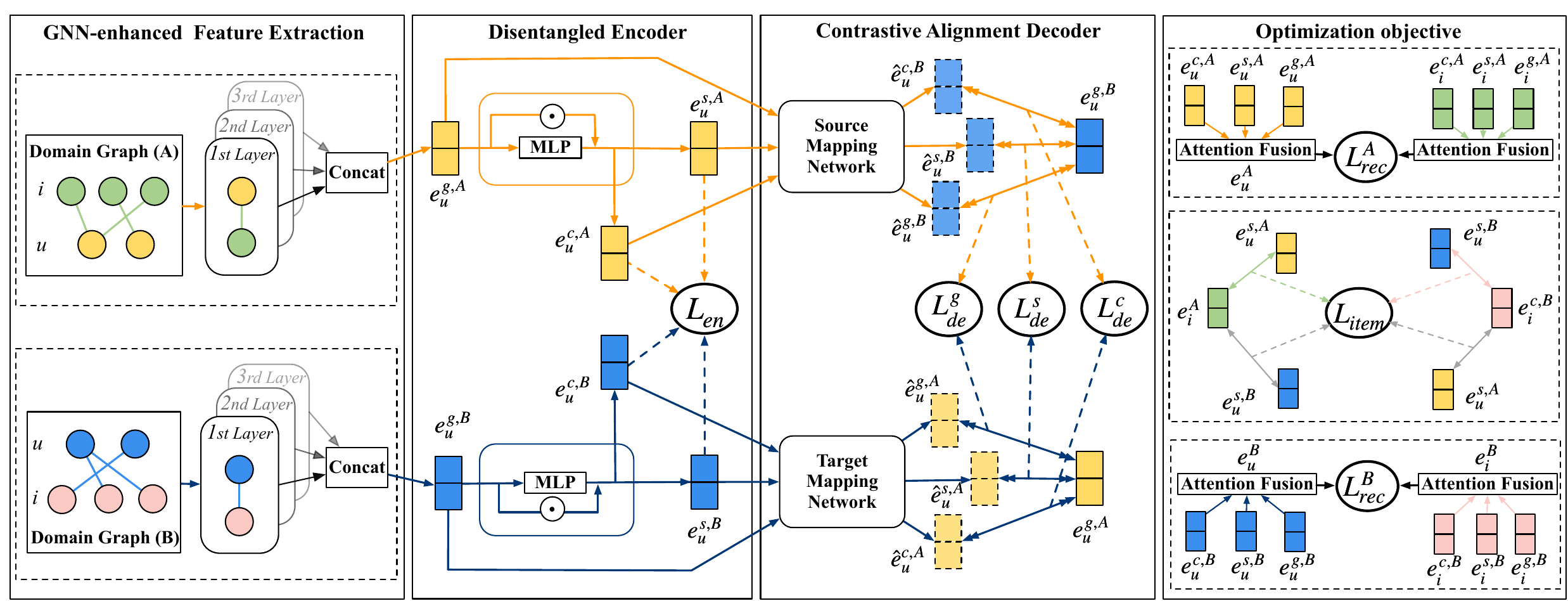}
		}
		\Description[Overview of the DGCDR framework]{This figure illustrates the overall framework of the proposed DGCDR model. From left to right, it consists of four parts. The first three parts represent the core modules of the model: (1) a GNN-enhanced feature extraction module that captures high-order collaborative signals in two domains, (2) a disentanglement encoder that separates domain-shared and domain-specific representations, and (3) a contrastive alignment decoder that incorporates explicit supervisory objectives by leveraging hierarchical contrastive loss. Each module is interconnected to reflect the flow of representation learning. The fourth part on the right presents three sub-figures that depict the model’s training objectives, corresponding to the three sub-loss components used for optimization.}
		\caption{Overview of the proposed DGCDR.}
		\label{fig:framework}
	\end{figure*}
	The overview of the proposed DGCDR is shown in Fig. \ref{fig:framework}. The model contains three main parts: GNN-enhanced feature extraction, disentanglement encoder, and contrastive alignment decoder. We will provide detailed descriptions in the subsequent subsections.
	\subsection{GNN-enhanced Feature Extraction}
	User and item representation learning is the foundation of recommender system. User history interactions serve as valuable information in the system and intuitively reflect user preferences. A user-item bipartite graph can be constructed by treating users and items as nodes and creating edges when interactions occur between nodes. We build two such heterogeneous graphs based on the interaction data of the $D_A$ and $D_B$ domains, respectively. Since graph convolutional networks (GCN) have shown powerful capability to capture collaborative signals in high-order connectivity \cite{GaoZLLQPQCJHL23}, our model employs GCN to learn these relationships within single-domain graphs. The intuition behind GCN-based graph modeling is that node features could be integrated by iteratively transforming, propagating, and aggregating features from their local neighborhood. We adopt collaborative filtering-based feature propagation \cite{LiuLLP20}, which is expressed as follows:
	\begin{equation}
		\begin{aligned}
			& \boldsymbol{e}_u^{(h+1)}=\boldsymbol{e}_u^{(h)}+\sum_{i \in \mathcal{N}_u} \frac{1}{\sqrt{\left|\mathcal{N}_u\right|\left|\mathcal{N}_i\right|}}(\boldsymbol{e}_i^{(h)}+\boldsymbol{e}_i^{(h)} \odot \boldsymbol{e}_u^{(h)}) ,\\
			& \boldsymbol{e}_i^{(h+1)}=\boldsymbol{e}_i^{(h)}+\sum_{u \in \mathcal{N}_i} \frac{1}{\sqrt{\left|\mathcal{N}_u\right|\left|\mathcal{N}_i\right|}}(\boldsymbol{e}_u^{(h)}+\boldsymbol{e}_u^{(h)} \odot \boldsymbol{e}_i^{(h)})
		\end{aligned}
	\end{equation}
	where $\odot$ denotes the element-wise product, ${e}_{u}^{(h)}$ and ${e}_{i}^{(h)}$ denote the learned embeddings of user $u$ and item $i$ after $h$-layer information propagation. $\mathcal{N}_u (\mathcal{N}_i)$ are the first-hop neighbors of $u$ ($i$), and the normalization term $\frac{1}{\sqrt{\left|\mathcal{N}_u\right|\left|\mathcal{N}_i\right|}}$ is used to avoid a sharp increase in the scale of embedding with propagation operations.
	
	After performing $H$-layer graph convolution, we concatenate all layer vectors $\{\boldsymbol{e}_{u/i}^{(0)},\dots, \boldsymbol{e}_{u/i}^{(H)}\}$ to generate embeddings that contain multi-order connectivity. We perform the same enrichment process for both $D_A$ and $D_B$ domains. The GNN-enhanced embeddings of user and item in $D_A$ are $\boldsymbol{e}_{u}^{g,A}=\boldsymbol{e}_{u^A}^{(0)}\|\cdots\| \boldsymbol{e}_{u^A}^{(H)}$, $\boldsymbol{e}_{i}^{g,A}=\boldsymbol{e}_{i^A}^{(0)}\|\cdots\| \boldsymbol{e}_{i^A}^{(H)}$. Analogously, we can derive $\boldsymbol{e}_{u}^{g,B}$ and $\boldsymbol{e}_{i}^{g,B}$ for $D_B$.
	\subsection{Disentangled Representation Encoder}
	The basic assumption of CDR is that users share several common interests among various domains, which are derived from the personalities of the users \cite{LiuLLP20, ZhuTLZXZLH22}. In addition to common interests, users have distinct preferences specific to different domains \cite{ZhangCLYP24}. Since GNN embeddings are abstract and condensed, it is necessary to distinguish the highly entangled domain-shared and domain-specific features to understand the complexity of user preference deeply. For a user $u$ in $D_A$ or $D_B$, we first map the GNN embedding into separate latent spaces, generating two independent representations for general and specific characteristics of $u$ respectively:
	\begin{equation}
		\begin{aligned}
			\boldsymbol{e}_u^{c, A}&=\boldsymbol{e}_u^{g, A} \odot MLP(\boldsymbol{e}_u^{g, A} ; \Theta_A^c), &\quad
			\hspace{-10pt} \boldsymbol{e}_u^{s, A}&=\boldsymbol{e}_u^{g, A} \odot MLP(\boldsymbol{e}_u^{g, A} ; \Theta_A^s)  \\
			\boldsymbol{e}_u^{c, B}&=\boldsymbol{e}_u^{g, B} \odot MLP(\boldsymbol{e}_u^{g, B} ; \Theta_B^c), &\quad
			\hspace{-10pt} \boldsymbol{e}_u^{s, B}&=\boldsymbol{e}_u^{g, B} \odot MLP(\boldsymbol{e}_u^{g, B} ; \Theta_B^s) 
		\end{aligned}
		\label{eq:projection}
	\end{equation}
	where superscript $c$ denotes the domain-shared latent space and $s$ denotes the domain-specific latent space. $\Theta_A$ and $\Theta_B$ are parameters of MLPs with activation function \textit{sigmoid}, respectively. Note that item features are also separated into domain-shared and -specific components ($\boldsymbol{e}_i^{c, A}$, $\boldsymbol{e}_i^{s, A}$, $\boldsymbol{e}_i^{c, B}$, $\boldsymbol{e}_i^{s, B}$) to align with user embedding spaces. This symmetry design ensures structural consistency and supports effective feature interactions in the recommendations.
	
	To encode user preferences in each domain into two distinct representations, we introduce constraints to ensure that the domain-shared and domain-specific representations distinctly reflect complementary aspects of user preferences.  For domain-shared representations, we align features across domains by minimizing their correlation discrepancy using a similarity measure. For domain-specific representations, we enforce their orthogonality with domain-shared features to minimize redundancy and ensure distinctiveness. These constraints enable the disentanglement process to focus on capturing transferable and domain-specific characteristics critical to CDR.  The disentangled encoder loss is formulated as follows:
	\begin{equation}
		\begin{aligned}
			\mathcal{L}_{e n}=dis(\boldsymbol{e}_u^{c, A}, \boldsymbol{e}_u^{c, B})+\|\boldsymbol{e}_u^{c, A}\cdot\boldsymbol{e}_u^{s, A}\|^2+\|\boldsymbol{e}_u^{c, B}\cdot\boldsymbol{e}_u^{s, B}\|^2+\Theta_{en}
		\end{aligned}
		\label{eq:encoderLoss}
	\end{equation}
	where $\Theta_{en}=\{\Theta_{en}^A, \Theta_{en}^B\}$, represents the learnable parameters. $dis(\cdot)$ measures the discrepancy between vectors, and we adopt the cosine distance as $dis(h_1, h_2)=1-\frac{h_1 \cdot h_2}{\|h_1\| \|h_2\|}$. 
	
	\noindent \textbf{Choice of Design.} It is worth noting that we also experimented with alternative strategies for disentanglement. Specifically, we experimented with a soft separation strategy by replacing the orthogonality constraint between domain-specific and domain-shared representations with a cosine similarity objective to encourage low correlation. However, this approach consistently led to inferior performance. Our analysis suggests that such soft constraints are insufficient to enforce semantic independence between intra-domain features. In contrast, orthogonality provides a clearer separation between subspaces, resulting in more robust representation disentanglement. As for the alignment of domain-shared representations across domains, we deliberately adopted a cosine-based distance instead of a stricter L2 objective, as it led to more stable training and improved downstream performance. These design decisions reflect our effort to customize the disentanglement process based on the specific roles of different representation types.
	
	Through the disentangled encoder, we can obtain the disentangled domain-shared and domain-specific features for users and items.  We then fuse the disentangled features and reconstruct user preferences in a personalized way. Some methods \cite{ZhangZZWWY23} adopt direct concatenation to fuse features. However, user preferences for intra-domain features can differ significantly in real-world applications \cite{LiuSNJ024}.  For instance, in the cross-domain scenario of movies and books, some users may prioritize domain-shared attributes like category and story topic, while others may focus on domain-specific attributes such as preferred directors or authors. This diversity underscores the importance of tailoring the feature fusion to individual user preferences. Therefore, we employ an attention mechanism to dynamically calculate the weights assigned to the disentangled feature in $D_A$ as follows (similarly for $D_B$):
	\begin{equation}
		[a_*^{c,A}, a_*^{s,A}] = \mathrm{softmax} \left( \frac{\boldsymbol{e}_*^{g,A} [\boldsymbol{e}_*^{c,A}, \boldsymbol{e}_*^{s,A}]}{\sqrt{d}} \right),\ * \in \{u, i\}
		\label{eq:atten}
	\end{equation}
	where $d$ represents the hidden dimension. We then fuse all representations according to these weights to obtain the final refined user and item embeddings:
	\begin{equation}
		\begin{aligned}
			\boldsymbol{e}_*^A=\boldsymbol{e}_*^{g, A}+a_*^{c, A} \boldsymbol{e}_*^{c, A}+a_*^{s, A} \boldsymbol{e}_*^{s, A}, \ * \in \{u, i\}
		\end{aligned}
		\label{eq:fusion}
	\end{equation}
	
	In order to achieve a more nuanced and precise disentanglement of user features, we also refine the user representation from the item view. Liu et al.\cite{LiuSNJ024} claimed that completely separating intra-domain features makes it infeasible to utilize a user's cross-domain specific features for item recommendation. Inspired by this, we intuitively leverage the item embedding to contrast the domain-specific features of the user and introduce an item contrastive loss $\mathcal{L}_{item}=\mathcal{L}_{item}^A$+$\mathcal{L}_{item}^B$, where $\mathcal{L}_{item}^A$ is defined as follows (with $\mathcal{L}_{item}^B$ being similar):
	\begin{equation}
		\begin{aligned}
			\mathcal{L}_{item}^A=\sum_{\left\langle u, i \right\rangle \in R^{A+}}-\log \frac{\exp(f(\boldsymbol{e}_u^{s, A}, \boldsymbol{e}_i^{A})/ \tau)}{\exp (f(\boldsymbol{e}_u^{s, A}, \boldsymbol{e}_i^{A})/ \tau)+\exp (f(\boldsymbol{e}_u^{s, B}, \boldsymbol{e}_i^{A})/ \tau)}
		\end{aligned}
		\label{eq: itemrLoss}
	\end{equation}
	where $R^{A+}$ denotes the interactions observed in $D_A$. $f(\cdot, \cdot)$ quantifies the mutual information of features by a dot product, and $\tau$ represents a temperature parameter. $\mathcal{L}_{item}^A$ ensures that the similarity between an item in $D_A$ and a user domain-specific feature in $D_A$ exceeds that of $D_B$, so as to reinforce the distinction between intra-domain features in respective domains.
	
	\subsection{Contrastive Alignment Decoder}
	Although the disentangled encoder forces representations to capture different aspects of user preference, the unsupervised learning process limits the model's ability to guarantee that each representation captures the desired properties (e.g., the domain-shared representation may not accurately capture the common characteristics). As demonstrated by Locatello et al. \cite{LocatelloBLRGSB19}, unsupervised disentanglement is highly sensitive to randomness and hyperparameters. This highlights the importance of incorporating task-specific supervisory signals to guide feature disentanglement and ensure robust representation learning. Without these signals, the disentanglement of the encoder may result in critical misclassifications, such as misclassifying domain-shared features as domain-specific or vice versa. Such misclassification can affect the model, especially when significant discrepancies exist between domains.
	
	However, constructing explicit supervision for disentangled representations in CDR is inherently challenging because domain-shared and domain-specific features are abstract and lack clear annotation boundaries. Even if feasible, the cost would be prohibitively high. To address this challenge, we draw inspiration from a modality-disentangled method \cite{Han0NL22} and propose a decoder that incorporates explicit supervisory objectives by leveraging hierarchical contrastive loss as an effective alternative. The decoder is able to guide the disentanglement process through an anchor mechanism, hence ensuring the alignment of domain-shared features. Specifically, we introduce an MLP-based mapping network for cross-domain knowledge transfer to transform domain-shared features:
	\begin{equation}
		\hat{\boldsymbol{e}}_u^{c, A}=\boldsymbol{e}_u^{c, B} \odot MLP(\boldsymbol{e}_u^{c, B} ; \Phi_B), \quad
		\hat{\boldsymbol{e}}_u^{c, B}=\boldsymbol{e}_u^{c, A} \odot MLP(\boldsymbol{e}_u^{c, A} ; \Phi_A)
		\label{eq:mapping}
	\end{equation}
	where $\hat{\boldsymbol{e}}_u^{c, A}$ and $\hat{\boldsymbol{e}}_u^{c, B}$ denote the domain-shared features transformed via $\boldsymbol{e}_u^{c, B}$ and $\boldsymbol{e}_u^{c, A}$ of the other domain, respectively. $\Phi_A$ and $\Phi_B$ are learnable parameters of mapping networks in two domains.
	
	To further stabilize the alignment and promote intra-domain separation under supervision, we also apply the same mapping network to GNN-enhanced (${\boldsymbol{e}}_u^{g, *}$) and domain-specific features (${\boldsymbol{e}}_u^{s, *}$). Specifically, similar to Eq. (\ref{eq:mapping}), we can obtain their transformed representations as follows:
	\begin{equation}
		\begin{aligned}
			\hat{\boldsymbol{e}}_u^{g, A}&=\boldsymbol{e}_u^{g, B} \odot MLP(\boldsymbol{e}_u^{g, B} ; \Phi_B),&\quad \hspace{-10pt}
			\hat{\boldsymbol{e}}_u^{g, B}&=\boldsymbol{e}_u^{g, A} \odot MLP(\boldsymbol{e}_u^{g, A} ; \Phi_A)\\
			\hat{\boldsymbol{e}}_u^{s, A}&=\boldsymbol{e}_u^{s, B} \odot MLP(\boldsymbol{e}_u^{s, B} ; \Phi_B),&\quad \hspace{-10pt}
			\hat{\boldsymbol{e}}_u^{s, B}&=\boldsymbol{e}_u^{s, A} \odot MLP(\boldsymbol{e}_u^{s, A} ; \Phi_A)
		\end{aligned}
	\end{equation}
	
	Since the mapping network is specifically designed to facilitate the transformation of transferable features, it focuses on domain-shared features that align with the mapping objectives. Therefore, features such as domain-specific representations should be suppressed by the network, as they are irrelevant to the desired mapping transformation. Taking inspiration from the modality-disentangled design in Han et al.\cite{Han0NL22}, we initially formulated this hierarchical relationship using a log-sigmoid of difference terms to reflect pairwise ordering between features. However, we found that this approach was unstable during training, possibly due to gradient saturation. To address this, we adopt a ratio-based contrastive loss to introduce more stable and effective task-specific supervision. Specifically, we use GNN-enhanced representations $\boldsymbol{e}_{u}^{g,A}$ and $\boldsymbol{e}_{u}^{g,B}$ as anchors, and compare their normalized similarity scores with those of the transformed domain-shared, domain-specific, and GNN-enhanced features from the other domain. This formulation follows the InfoNCE loss framework \cite{abs-1807-03748}, where each anchor is encouraged to be more similar to its positive sample than to the negative one, thereby establishing a supervised hierarchical ordering of representations. Two types of loss are driven by the hierarchical alignment relationships: (1) domain-shared features should be more aligned with the anchor than GNN-enhanced features, denoted as $\mathcal{L}_{de, c \to g}$, and (2) GNN-enhanced features should be more aligned than domain-specific features, denoted as $\mathcal{L}_{de, g \to s}$. While traditional InfoNCE considers a single positive and multiple negatives in a batch-wise setting, we adopt a pairwise variant of contrastive comparison, computing local two-way ranking losses that reflect hierarchical pairwise preference relations between disentangled representations. The following formulation is for domain $D_A$, where $\boldsymbol{e}_{u}^{g,A}$ serves as the anchor, and the transformed features originate from $D_B$:
	\begin{equation}
		\begin{aligned}
			\mathcal{L}_{de, c \to g}^{A} & =-\log \frac{\exp (f(\boldsymbol{e}_u^{g, A}, \hat{\boldsymbol{e}}_u^{c, A}) / \tau)}{\exp (f(\boldsymbol{e}_u^{g, A}, \hat{\boldsymbol{e}}_u^{c, A}) / \tau)+\exp (f(\boldsymbol{e}_u^{g, A}, \hat{\boldsymbol{e}}_u^{g, A}) / \tau)}, \\
			\mathcal{L}_{de, g \to s}^{A} & =-\log \frac{\exp (f(\boldsymbol{e}_u^{g, A}, \hat{\boldsymbol{e}}_u^{g, A}) / \tau)}{\exp (f(\boldsymbol{e}_u^{g, A}, \hat{\boldsymbol{e}}_u^{g, A}) / \tau)+\exp (f(\boldsymbol{e}_u^{g, A}, \hat{\boldsymbol{e}}_u^{s, A}) / \tau)}
		\end{aligned}
	\end{equation}
	where $f(\cdot, \cdot)$ denotes the dot-product similarity between feature pairs, and $\tau$ is a temperature parameter. A symmetric formulation is applied to domain $D_B$, where $\boldsymbol{e}_{u}^{g,B}$ serves as the anchor. 
	
	The total decoder loss sums the contrastive objectives across both domains:
	\begin{equation}
		\mathcal{L}_{de} = \mathcal{L}_{de, c \to g}^{A} + \mathcal{L}_{de, g \to s}^{A} + \mathcal{L}_{de, c \to g}^{B} + \mathcal{L}_{de, g \to s}^{B}
		\label{eq:decoderLoss}
	\end{equation}
	This contrastive objective provides explicit supervision for preserving hierarchical relationships among disentangled representations and improves gradient smoothness and convergence stability.
	
	\subsection{Prediction and Optimization objective}
	Finally, we obtain the predicted scores via the inner product of the final user and item embeddings from Eq. (\ref{eq:fusion}) in $D_A$, $\hat{r}^A_{u,i}=\boldsymbol{e}_u^A\cdot\boldsymbol{e}_i^A$, similarly to domain $D_B$, $\hat{r}^B_{u,i}=\boldsymbol{e}_u^B\cdot\boldsymbol{e}_i^B$.
	
	Following previous work \cite{LiuSNJ024, Song0DZWB024, XuXCZ21, GuoH2025}, we employ Bayesian Personalized Ranking (BPR) loss to optimize the recommendation of the model,  $\mathcal{L}_{rec}=\mathcal{L}_{rec}^A+\mathcal{L}_{rec}^B$, where $\mathcal{L}_{rec}^A$ is as follows:
	\begin{equation}
		\mathcal{L}_{rec}^A= -\sum_{\left\langle u, i^{+}, i^{-}\right\rangle \in \mathcal{T}_{d}^A}\ln \sigma\left(\hat{r}_{u, i^{+}} - \hat{r}_{u, i^{-}}\right)
	\end{equation}
	where $\sigma$ is \textit{sigmoid} function, and the triplet set $\mathcal{T}_{d}^A=\{\langle u, i^{+}, i^{-}\rangle\}$ represents a training mini-batch in $D_A$. During training, each positive item is paired with one sampled negative item in a uniform distribution. Since user $u$ expresses a preference for item $i^{+}$ over item $i^{-}$, BPR loss encourages assigning higher scores to observed user-item interactions than unobserved ones.
	
	Combining encoder loss defined in Eq. (\ref{eq:encoderLoss}), item contrastive loss in Eq. (\ref{eq: itemrLoss}) and decoder loss in Eq. (\ref{eq:decoderLoss}), the overall loss of DGCDR is:
	\begin{equation}
		\mathcal{L} = \mathcal{L}_{rec} + \lambda_{en}\mathcal{L}_{en} + \lambda_{de}\mathcal{L}_{de} + \lambda_{item}\mathcal{L}_{item} + \lambda\|\Theta\|_{2}^{2}
	\end{equation}
	where $\Theta$ is parameters involved in the model, regularized using \textit{L2}.
	
	\section{Experiments} \label{experiments}
	In this section, we conduct experiments to answer the following questions: 
	\begin{description}[leftmargin=1.8em]
		\item[Q1] How does DGCDR perform compared to SOTA competitors?
		\item[Q2] Do the key components of DGCDR effectively contribute to performance improvement?
		\item[Q3] How does DGCDR achieve the desired disentanglement?
		\item[Q4] Can our model evaluate the importance of intra-domain features in cross-domain recommendation?
	\end{description}
	
	\subsection{Datasets}
	We evaluate the proposed model and baselines on two real-world datasets, \textit{Amazon} \cite{hou2024bridging} and \textit{Douban} \cite{zhao2021recbole}, using three domain pairs, each serving as the target domain in turn, forming six cross-domain tasks. The datasets are described as follows: 
	\begin{itemize} 
		\item \textit{Amazon}\footnote{https://amazon-reviews-2023.github.io/index.html} is a large-scale e-commerce dataset with user-item interactions across multiple domains. We select two domain pairs: Cloth\&Sport and Cloth\&Electronics (abbreviated as Elec). Cloth and Sport are more closely related in domain, whereas Cloth and Elec share less knowledge. 
		\item \textit{Douban} is a media platform covering movies, books, and music. We construct two tasks by alternating movies and books as source and target domains. 
	\end{itemize}
	To ensure that the dataset satisfies the requirements of $N$-core \cite{SunY00Q0G20} and full user overlap for dual-target cross-domain scenarios\footnote{We adopt the full-overlap setting to highlight the effectiveness of disentanglement. Anchor-based supervision is only applied to users with interactions in both domains, while non-overlapping users can still benefit from single-domain GNN-based propagation. Our method also performs robustly under partial-overlap scenarios.}, we apply the following iterative sampling for each domain pair:
	\begin{enumerate}[leftmargin=3.1em]
		\item [\textit{Step 1.}] Extract overlapping users from both domains;
		\item [\textit{Step 2.}] Filter users: remove those with fewer than $N$ interactions;
		\item [\textit{Step 3.}] Filter items: remove those with fewer than $N$ interactions;
		\item [\textit{Step 4.}] Iterate: repeat the \textit{Steps 2} and \textit{3} until all users and items have at least $N$ interactions, since removing items may reduce user interactions to less than $N$, and vice versa for users;
		\item [\textit{Step 5.}] Re-extract overlapping users in both domains after filtering.
	\end{enumerate}
	We adopt commonly used values for the $N$-core filtering, following established practices \cite{ZhouL0M23, Han0NL22}. Specifically, we set $N$=10 for AmazonElec\&Cloth and DoubanMovie\&Book domain pairs, and $N$=5 for AmazonSport\&Cloth to ensure a sufficient number of eligible data during iterative sampling. The detailed statistics of the sampled datasets are presented in Table \ref{tab:dataset_stats}.
	\begin{table}[]
		\caption{Dataset statistics}
		\label{tab:dataset_stats}
		\centering
		\begin{tabular}{lcccc}
			\toprule
			\textit{Dataset} & \#\textit{OverlapUsers} & \#\textit{Items} & \#\textit{Inters} & \textit{Sparsity} \\
			\midrule
			AmazonElec & \multirow{2}{*}{35,827} & 62,548 & 811,969 & 99.9638\% \\
			AmazonCloth &  & 72,669 & 847,042 & 99.9675\% \\
			\midrule
			AmazonSport & \multirow{2}{*}{149,520} & 94,057 & 1,474,955 & 99.9895\% \\
			AmazonCloth &  & 149,490 & 2,222,238 & 99.9901\% \\
			\midrule
			DoubanMovie & \multirow{2}{*}{10,654} & 18,833 & 2,287,871 & 98.8598\% \\
			DoubanBook & & 16,014 & 636,812 & 99.6268\% \\
			\bottomrule
		\end{tabular}
	\end{table}
	\subsection{Experimental setup}
	\subsubsection{Evaluation protocol}
	In line with previous research, the data in the target domain is split into 60\% training, 20\% validation, and 20\% test sets, while the source domain data is divided into 80\% training and 20\% validation. Performance is evaluated on the target domain using four widely adopted metrics \cite{KangHLY19, ZhangZZWWY23, guo2024configurable}: Recall, Hit Ratio (HR), Mean Reciprocal Rank (MRR), and Normalized Discounted Cumulative Gain (NDCG). Recall and HR measure retrieval accuracy, while MRR and NDCG assess both the relevance and ranking of retrieved items. All results are averaged over five runs with different random seeds, and statistical significance is based on these repeated runs.
	
	\subsubsection{Competitors}
	We compare our proposed DGCDR with representative state-of-the-art methods from both single-domain and cross-domain recommendation settings, focusing on methods that use GNNs and apply feature disentanglement, which are also the main components of our approach.
	
	\textbf{Cross-domain methods} utilize both source and target domains for joint training. We include:
	(1) the pioneering dual-target model \textbf{DTCDR}\cite{ZhuC0LZ19},
	(2) GNN-based models such as \textbf{BiTGCF}\cite{LiuLLP20}, \textbf{DCCDR}\cite{ZhangZZWWY23}, and \textbf{DRLCDR}\cite{ZhangCLYP24}, and
	(3) disentanglement-based approaches including \textbf{DCCDR} and \textbf{DRLCDR}.
	These methods cover various strategies such as multi-task learning, bi-directional transfer, and contrastive regularization.
	\textbf{Single-domain methods} rely solely on target domain data. We consider GNN-based models like \textbf{LightGCN}\cite{0001DWLZ020}, \textbf{NCL}\cite{LinTHZ22}, and \textbf{DGCF}\cite{WangJZ0XC20}, with \textbf{DGCF} also modeling disentangled user intents. These baselines allow us to evaluate the contributions of GNN and disentanglement.
	\setlength{\tabcolsep}{6pt}
	
	\subsubsection{Implementation and hyperparameter setting}
	\begin{table*}[h]
		\footnotesize
		\setlength{\tabcolsep}{6pt}
		\caption{Performance comparison between baselines and our model. The best performance is in boldface, and the second is underlined. Symbol $\dagger$ after the name indicates a method based on disentanglement, and the symbol * denotes the statistical significance of our method compared to the optimal baselines (paired t-test, p-value\textless0.05).}
		\label{tab:OverallComp}
		\resizebox{0.83\linewidth}{!}{ 
			\begin{tabular}{@{}llccccccccc@{}}
				\toprule
				\multirow{3}{*}{\textbf{Dataset}} & \multirow{3}{*}{\textbf{Metric}} & \multicolumn{3}{c}{\multirow{2}{*}{\textbf{Single-domain Methods}}} & \multirow{3}{*}{} & \multicolumn{5}{c}{\multirow{2}{*}{\textbf{Cross-domain Methods}}}               \\
				&                                  & \multicolumn{3}{c}{}                                        &                   & \multicolumn{5}{c}{}                                                     \\ \cmidrule(lr){3-5} \cmidrule(l){7-11} 
				&                                  & LightGCN          & NCL               & DGCF$\dagger$       &                   & DTCDR & DCCDR$\dagger$ & DRLCDR$\dagger$ & BiTGCF      & \textbf{DGCDR}$\dagger$  \\ \midrule
				\multirow{4}{*}{Cloth}            & Recall                           & .0222             & .0232             & .0231               &                   & .0217 & .0217          & .0195           & \underline{.0233} & \textbf{.0260*} \\
				& MRR                              & .0227             & .0251             & .0243               &                   & .0244 & .0253          & \underline{.0276}     & .0254       & \textbf{.0278*} \\
				& HR                               & .0735             & \underline{.0795}       & .0786               &                   & .0713 & .0726          & .0678           & .0788       & \textbf{.0876*} \\
				& NDCG                             & .0144             & .0155             & .0153               &                   & .0150 & .0152          & .0151           & \underline{.0157} & \textbf{.0173*} \\
				\multirow{4}{*}{Elec}             & Recall                           & .0381             & \underline{.0395}       & .0374               &                   & .0347 & .0350          & .0339           & .0383       & \textbf{.0403*} \\
				& MRR                              & .0337             & .0339             & .0335               &                   & .0306 & .0312          & \textbf{.0354}  & .0335       & \underline{.0353}     \\
				& HR                               & .1225             & \underline{.1260}       & .1204               &                   & .1099 & .1119          & .1216           & .1241       & \textbf{.1304*} \\
				& NDCG                             & .0234             & \underline{.0240}       & .0232               &                   & .0216 & .0217          & .0223           & .0234       & \textbf{.0247*} \\ 
				\midrule
				\multirow{4}{*}{Cloth}            & Recall                           & .0193             & .0198             & .0193               &                   & .0181 & .0183          & .0176           & \underline{.0228} & \textbf{.0252*} \\
				& MRR                              & .0146             & .0155             & .0153               &                   & .0151 & .0147          & \textbf{.0200}  & .0168       & \underline{.0176}     \\
				& HR                               & .0404             & .0410             & .0404               &                   & .0367 & .0371          & .0440           & \underline{.0463} & \textbf{.0503*} \\
				& NDCG                             & .0115             & .0121             & .0119               &                   & .0117 & .0114          & .0128           & \underline{.0137} & \textbf{.0146*} \\
				\multirow{4}{*}{Sport}            & Recall                           & \underline{.0254}       & .0240             & .0243               &                   & .0174 & .0207          & .0158           & .0246       & \textbf{.0275*} \\
				& MRR                              & \underline{.0106}       & .0102             & .0104               &                   & .0061 & .0082          & .0071           & .0100       & \textbf{.0112*} \\
				& HR                               & \underline{.0412}       & .0396             & .0402               &                   & .0261 & .0331          & .0307           & .0402       & \textbf{.0446*} \\
				& NDCG                             & \underline{.0118}       & .0113             & .0114               &                   & .0075 & .0094          & .0073           & .0113       & \textbf{.0126*} \\ 
				\midrule
				\multirow{4}{*}{Book}             & Recall                           & \underline{.1356}       & .1316             & .1302               &                   & .0735 & .0650          & .0698           & .1314       & \textbf{.1369*} \\
				& MRR                              & .1476             & .1361             & .1431               &                   & .1044 & .0967          & .1068           & \underline{.1498} & \textbf{.1557*} \\
				& HR                               & \underline{.5394}       & .5185             & .5300               &                   & .3724 & .3357          & .3663           & .5329       & \textbf{.5505*} \\
				& NDCG                             & \underline{.0922}       & .0871             & .0888               &                   & .0552 & .0496          & .0543           & .0917       & \textbf{.0954*} \\
				\multirow{4}{*}{Movie}            & Recall                           & .0687             & \underline{.1027}       & .0987               &                   & .0728 & .0811          & .0746           & .0993       & \textbf{.1037*} \\
				& MRR                              & .1791             & .2050             & \underline{.2165}         &                   & .1833 & .1976          & .1783           & .1900       & \textbf{.2190*} \\
				& HR                               & .5723             & \underline{.6848}       & .6756               &                   & .5934 & .6269          & .5742           & .6621       & \textbf{.6907*} \\
				& NDCG                             & .0849             & .1065             & \underline{.1071}         &                   & .0876 & .0949          & .0861           & .0978       & \textbf{.1117*} \\ 
				\bottomrule
			\end{tabular}
	}\end{table*}
	Our model and most baselines are implemented in PyTorch using the RecBole library \cite{zhao2021recbole}, while DRLCDR leverages its original public code. All experiments are deployed on an Nvidia RTX 3090 GPU (24GB). To ensure fairness, we standardize the latent dimension to 64 in all methods. All layer parameters are initialized with the Xavier normal as provided by PyTorch. We use fine-tuning to select hyperparameters in the following parameter sets\footnote{We have published the final hyperparameters in our Github project.}: subloss weights (i.e., $\lambda_{en}$, $\lambda_{de}$ and $\lambda_{item}$) from $\{0.01, 0.1, 1\}$ (see Sec.\ref{sec: paraAnaly}), L2 regularization coefficient from $\{1e^{-3}, 1e^{-4}, 1e^{-5}\}$, dropout rate from $\{0.1, 0.2, 0.3\}$, temperature parameter from 0.05 to 0.3 with a step of 0.05, and learning rate from $\{1e^{-3}, 1e^{-4}\}$. Additionally, we fix the GCN depth $H=3$, as third-order information is able to represent user interests, behavioral similarity, and collaborative signals \cite{WangJZ0XC20}. The training batch size is set to 2048 for Sport\&Cloth and 4096 for other tasks, adjusted based on hardware capacity. The Adam optimizer is employed for parameter updates. We adopt early stopping with patience for 10 epochs to avoid overfitting. All baselines are fine-tuned using their original hyperparameters for optimal results.
	
	\subsection{Overall performance (Q1)}\label{sec:overallPerform}
	Table \ref{tab:OverallComp} presents the results of the performance comparison\footnote{The small metric values are due to efficiency-oriented settings, such as larger batch sizes and candidate sets.}, and we have the key observations in this experiment as follows:
	
	(1) \textit{For Our DGCDR}. The model consistently outperforms baselines across almost all metrics. It is noteworthy that DGCDR achieves improvements of up to 11.59\% in the low-correlation pair (Elec\&Cloth) and 10.53\% in the high-correlation domain pair (Sport\&Cloth). This highlights the robustness of the model across varying domain correlations. The performance gain is attributed to the design of a GNN-enhanced disentanglement framework with supervisory signals, which enables effective disentanglement of intra-domain features and alignment of domain-shared features, ensuring improved feature consistency and transferability. 
	
	(2) \textit{For cross-domain methods}. Disentangled baselines achieve competitive results in certain cases but lack stability across domains. We attribute the reason to their use of unsupervised disentanglement techniques.  For instance, DCCDR applies self-supervised contrastive learning for feature disentanglement but struggles to handle domain discrepancies effectively; DRLCDR follows the VAE framework to model the data distribution in an unsupervised manner. In addition, BiTGCF utilizes GNN to aggregate high-order collaborative information, achieving promising performance. However, it simply interpolates between user embeddings in the two domains, neglecting explicit disentanglement of user preferences. In contrast, our DGCDR outperforms these methods, achieving average improvements of 32.36\% over DCCDR, 33.48\% over DRLCDR, and 8.09\% over BiTGCF. These results emphasize the importance of explicitly decoupling user preferences and tailoring domain-specific information to improve cross-domain recommendations.  
	
	(3) \textit{For single-domain methods}. LightGCN and NCL perform competitively by utilizing advanced GNN architectures, while DGCF combines GNNs with disentanglement learning, showing improved performance in some domains.  However, DGCDR surpasses all three methods (15.43\%, 9.90\%, and 10.52\% on average, respectively), particularly in sparse domains such as Cloth, where collaborative signals are often difficult to extract. This highlights the effectiveness of disentangled cross-domain recommendations, especially in scenarios with limited data availability.
	
	(4) \textit{For training cost}. As shown in Table~\ref{tab:trainCost}, the main overhead of DGCDR stems from the disentanglement and GNN-enhanced modules. Despite this, DGCDR maintains a favorable balance between efficiency and performance. Notably, compared to other disentangled baselines (marked with $\dagger$), DGCDR achieves consistently better accuracy with comparable or even reduced training time, highlighting its effective design. This demonstrates that the introduction of supervision and GNNs does not significantly compromise training efficiency, while enhancing model effectiveness.
	\begin{table}[]
		\footnotesize
		\setlength{\tabcolsep}{2pt} 
		\caption{Training time of cross-domain methods (seconds)}
		\centering
		\resizebox{1\linewidth}{!}{ 
			\begin{tabular}{@{}lccccccccc@{}}
				\toprule
				\textit{Dataset} & LightGCN          & NCL           & DGCF† & DTCDR & DCCDR† & DRCDR† & BiTGCF & DGCDR† \\
				\midrule
				Cloth & 943 & 7,993   & 10,172 &  5,736  & 1,183  & 5,697   & 7,350   & 9,192   \\
				Elec    & 4,205 & 3,032 & 51,336& 15,684 & 6,357  & 4,576   & 7,514   & 9,756   \\ \midrule
				Cloth    & 33,761  & 7,873   & 69,544&47,265 & 14,184 & 39,262  & 58,290  & 34,319  \\
				Sport    & 24,978  & 9,525  & 73,936 & 10,791 & 232,978& 17,560  & 28,677  & 13,140  \\ \midrule
				Book   & 5,839   & 1,336 & 31,883& 3,337  & 620   & 1,540   & 4,169   & 4,125   \\
				Movie & 201  & 8,451  & 51,910& 762   & 3,601  & 2,749   & 2,309   & 6,724   \\ \bottomrule
		\end{tabular}}
		\label{tab:trainCost}
	\end{table}

	\subsection{Discussion of Model Variants (Q2)}\label{sec:ablation}
	We conduct an ablation study to evaluate the contribution of each key component in DGCDR to overall performance. The following variants are implemented for comparison:
	\begin{itemize}
		\item {\textit{GCN}}: The basic variant removes other components and directly predicts using GCN embedding with a total loss of $\mathcal{L} = \mathcal{L}_{rec}$.
		\item {\textit{-Dec}}: removes the contrastive decoder loss in Eq. (\ref{eq:decoderLoss}) and the total loss $\mathcal{L} = \mathcal{L}_{rec}+ \lambda_{en}\mathcal{L}_{en}+ \lambda_{item}\mathcal{L}_{item}$.
		\item {\textit{-Enc}}: removes the disentangled encoder loss in Eq. (\ref{eq:encoderLoss}) and the total loss $\mathcal{L} = \mathcal{L}_{rec}+ \lambda_{de}\mathcal{L}_{de}+ \lambda_{item}\mathcal{L}_{item}$.
		\item {\textit{-ItDis}}: removes the item disentanglement loss in Eq. (\ref{eq: itemrLoss}) and the total loss $\mathcal{L} = \mathcal{L}_{rec}+ \lambda_{en}\mathcal{L}_{en}+ \lambda_{de}\mathcal{L}_{de}$.
		\item {\textit{-Pers}}: replaces personalized fusion in Eq. (\ref{eq:fusion}) with direct concatenation of features.
		\item{\textit{MarDec}}: replaces the original ratio-based hierarchical contrastive loss in the decoder (Eq. (\ref{eq:decoderLoss})) with margin-based losses.
	\end{itemize}
	\begin{table}[]
		\setlength{\tabcolsep}{3pt}
		\caption{Ablation results on key component}
		\label{tab:AblationRes}
		\begin{tabular}{@{}ccccccccc@{}}
			\toprule
			\multicolumn{2}{c}{Dataset}                          & \textit{GCN} & \textit{-Dec} & \textit{-Enc} & \textit{-ItDis} & \textit{-Pers} & \textit{MarDec} & DGCDR          \\ \midrule
			\multirow{2}{*}{Cloth} & \multicolumn{1}{c|}{Recall} & .0240        & .0251         & .0246         & .0250           & .0223          &.0253		& \textbf{.0260} \\
			& \multicolumn{1}{c|}{HR}     & .0828        & .0857         & .0838         & .0850           & .0742          &.0866	& \textbf{.0876} \\
			\multirow{2}{*}{Elec}  & \multicolumn{1}{c|}{Recall} & .0394        & .0397         & .0399         & .0401           & .0349          &.0397		& \textbf{.0403} \\
			& \multicolumn{1}{c|}{HR}     & .1271        & .1281         & .1284         & .1293           & .1087          &.1285		& \textbf{.1304} \\ \midrule
			\multirow{2}{*}{Cloth} & \multicolumn{1}{c|}{Recall} & .0203        & .0231         & .0241         & .0239           & .0196          &.0229		& \textbf{.0252} \\
			& \multicolumn{1}{c|}{HR}     & .0422        & .0474         & .0488         & .0486           & .0391          &.0470	& \textbf{.0503} \\
			\multirow{2}{*}{Sport} & \multicolumn{1}{c|}{Recall} & .0268        & .0267         & .0267         & .0267           & .0206          &.0266		& \textbf{.0275} \\
			& \multicolumn{1}{c|}{HR}     & .0435        & .0438         & .0434         & .0434           & .0331          &.0431	& \textbf{.0446} \\ \midrule
			\multirow{2}{*}{Book}  & \multicolumn{1}{c|}{Recall} & .1351        & .1367         & .1337         & .1356           & .0928          &.1358		& \textbf{.1369} \\
			& \multicolumn{1}{c|}{HR}     & .5451        & .5504         & .5384         & .5452           & .4304          &.5473		& \textbf{.5505} \\
			\multirow{2}{*}{Movie} & \multicolumn{1}{c|}{Recall} & .0999        & .1020         & .1033         & .1026           & .1032          &.1029		& \textbf{.1037} \\
			& \multicolumn{1}{c|}{HR}     & .6804        & .6843         & .6885         & .6867           & .6850          &.6903	& \textbf{.6907} \\ \bottomrule
		\end{tabular}
	\end{table}
	The results on Table \ref{tab:AblationRes} show that: 
	(1) Removing the personalized fusion mechanism leads to the most significant performance drop (22.69\% on average), highlighting its role in effectively integrating domain-shared and -specific features by tailoring fusion to individual preferences. 
	(2) The variant \textit{GCN} performs poorly (6.26\% drop), suggesting that relying solely on GCN embeddings is insufficient to capture users' complex domain preferences, further emphasizing the necessity of disentanglement. 
	(3) Removal of encoder and decoder loss leads to performance declines of 2.63\% and 2.66\%, respectively, demonstrating their importance in feature separation and alignment. 
	(4) The variant \textit{MarDec} consistently underperforms across all evaluation metrics, with an average performance drop of 2.76\%, and generally requires more epochs to converge.
	(5) The effect of item disentanglement loss is also meaningful as it enables the refinement of interaction-level specificity. These results validate the effectiveness of each component in DGCDR for enhancing cross-domain preference learning.
	
	\subsection{Analysis of Disentanglement and Alignment (Q3)}\label{sec:analysisDisen}
	Our DGCDR model employs a disentangled encoder for cross-domain feature matching and intra-domain feature separation, and a contrastive decoder for transferable feature alignment and disentanglement enhancement. To further validate this design,  we introduce two model variants: \textit{DGCDR w/o enLoss}, which excludes encoder loss $\mathcal{L}_{en}$ ($\lambda_{en}=0$), and \textit{DGCDR w/o en\&deLoss}, which excludes both encoder loss $\mathcal{L}_{en}$ and decoder loss $\mathcal{L}_{de}$ ($\lambda_{en}=0, \lambda_{de}=0$). We use t-SNE \cite{van2008visualizing} to visualize intra-domain features of these variants\footnote{The perplexity of t-SNE is set to 100, and the learning rate is fixed to 10, and we use the default settings provided by Scikit-learn for the other parameters.}. For faster results, we sample the data down to 10,000 for t-SNE training \cite{SmilkovTNRVW16}; for better visualization, we plot 500 randomly selected points for each feature type to avoid excessive overlap. As shown in Fig. \ref{fig:visualizeDisen}, we color-code the intra-domain features and make the following observations:

	\begin{figure}[]
		\includegraphics[width=\linewidth]{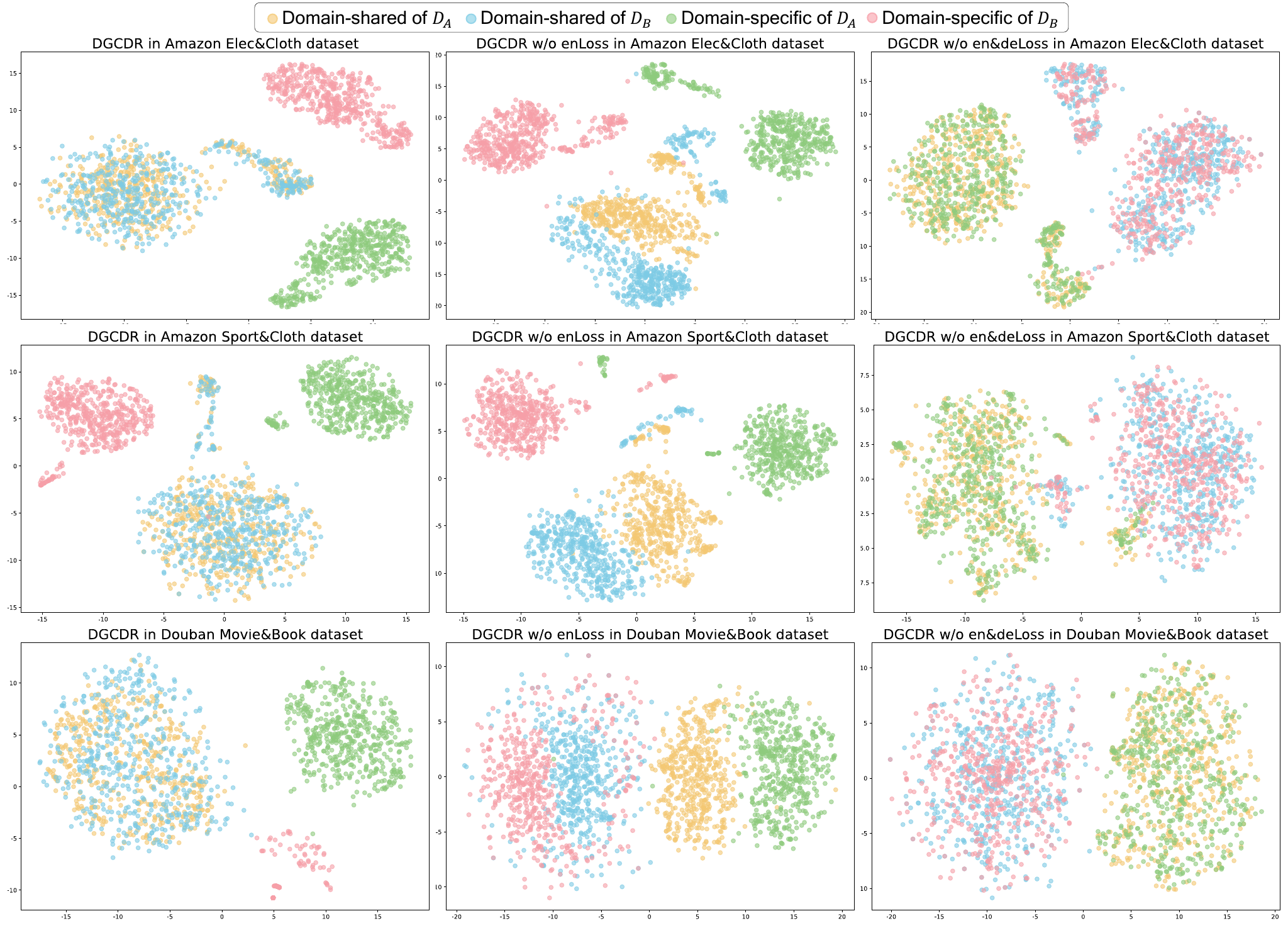}
		\Description[Effectiveness of DGCDR variants on feature disentanglement and alignment]{Each row in the figure visualizes the disentangled intra-domain features for a specific dataset using t-SNE, while each column compares different variants of the DGCDR model. The first column shows that the full DGCDR achieves both cross-domain alignment and intra-domain disentanglement. The second column (DGCDR without encoder loss) shows weakened domain-shared alignment. The third column (without both encoder and decoder losses) shows severe entanglement between domain-shared and domain-specific features, confirming the effectiveness of the proposed losses.}
		\caption{The t-SNE visualization of the disentangled intra-domain features, where each row shows three variants on the same dataset, and each column shows the effect of a variant across different datasets.}
		\label{fig:visualizeDisen}
	\end{figure}

	(1) The first column presents the effect of the complete DGCDR, where domain-shared features from two domains (i.e., blue and yellow points) have a high degree of overlap and exhibit similar manifold distributions, thus demonstrating cross-domain feature alignment. In addition, domain-shared and -specific features of the same domain (i.e., yellow vs. green, blue vs. pink) have no overlap, thus validating intra-domain feature disentanglement.
	
	(2) The effect of \textit{DGCDR w/o enLoss} is shown in the second column, where the overlap of the two domain-shared features is considerably reduced due to the absence of constraints on cross-domain matching for common knowledge. When we further remove the decoder constraints, \textit{DGCDR w/o en\&deLoss} exhibits the pattern of complete overlap of intra-domain features, indicating highly entangled domain-shared and -specific features in the same domain. Therefore, this experiment demonstrates the effectiveness of our modular design for cross-domain feature alignment and domain feature disentanglement.
	
	\subsection{Effect of Disentangled Features (Q4)}\label{sec: attenVisual}
	In this section, we analyze how different types of domain features influence performance across datasets, aiming to identify the most beneficial disentangled features during the cross-domain recommendation stage. Specifically, we reload the optimized model to visualize the average user attention scores derived from Eq. (\ref{eq:atten}), where higher attention scores indicate more valuable features for personalized fusion. The resulting distribution across the three datasets is shown in Fig.\ref{fig:attention}, and we have the following observations: 

	\begin{figure}[]
		\centering
		\includegraphics[width=\linewidth]{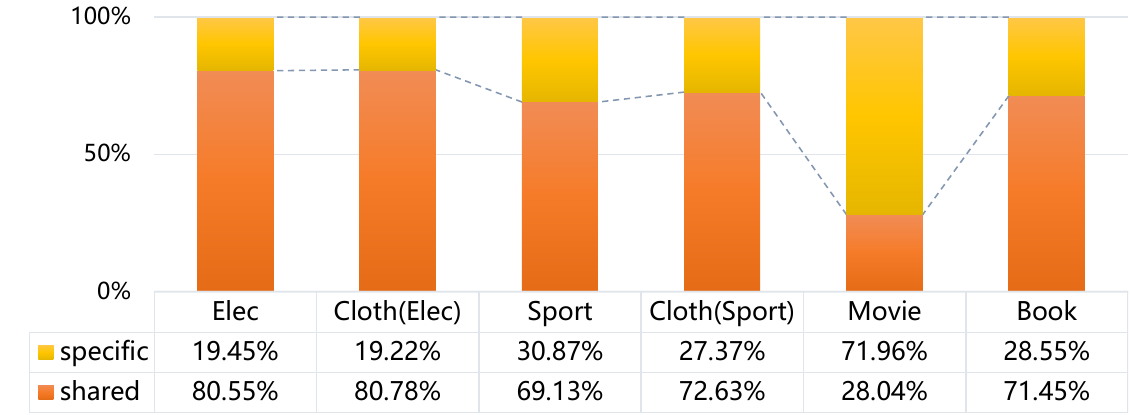}
		\Description[Attention weights on domain-shared and domain-specific features across six domains]{Bar chart showing the average attention weights of users to domain-shared and domain-specific features across six domains. In less related domains such as Electronics (80.55\% shared, 19.45\% specific) and Clothing (80.78\%, 19.22\%), higher emphasis is placed on domain-shared features. In more related or personalized domains such as Sports (69.13\%, 30.87\%) and Clothing (72.63\%, 27.37\%), attention shifts towards domain-specific features. Notably, Movie (28.04\%, 71.96\%) exhibits strong preference for specific features, while Book (71.45\%, 28.55\%) favors shared features.}
		\caption{Attention distribution of intra-domain features}
		\label{fig:attention}
	\end{figure}
	
	(1) \textit{From the perspective of Domain Relevance}, higher weights are assigned to domain-shared features in less related domains (i.e., Electronics and Clothing) to alleviate domain discrepancies. In contrast, closely related domains (i.e., Sports and Clothing) tend to focus more on domain-specific features to enhance personalization.
	
	(2) \textit{From the perspective of Domain Type}, content consumption domains (i.e., Movies and Books on Douban) exhibit different patterns compared to Amazon e-commerce domains. Movie domain favors domain-specific features due to higher personalization demands, while the Book domain prioritizes domain-shared features. This could be due to data sparsity, where leveraging domain-shared features from cross-domain learning can provide higher robustness in recommendations.
	
	\subsection{Hyperparameter Analysis}\label{sec: paraAnaly}
	We analyze two key hyperparameters in our model, $\lambda_{en}$ and  $\lambda_{de}$, which control the weights of encoder and decoder losses. Given the potential mutual influence between the two parameters, we chose a 2D heat map rather than separate analyses. Both parameters are varied across the range $\{0.01, 0.1, 1.0\}$ to jointly assess performance.
	
\begin{figure}[]
		\centering
		\includegraphics[width=\linewidth]{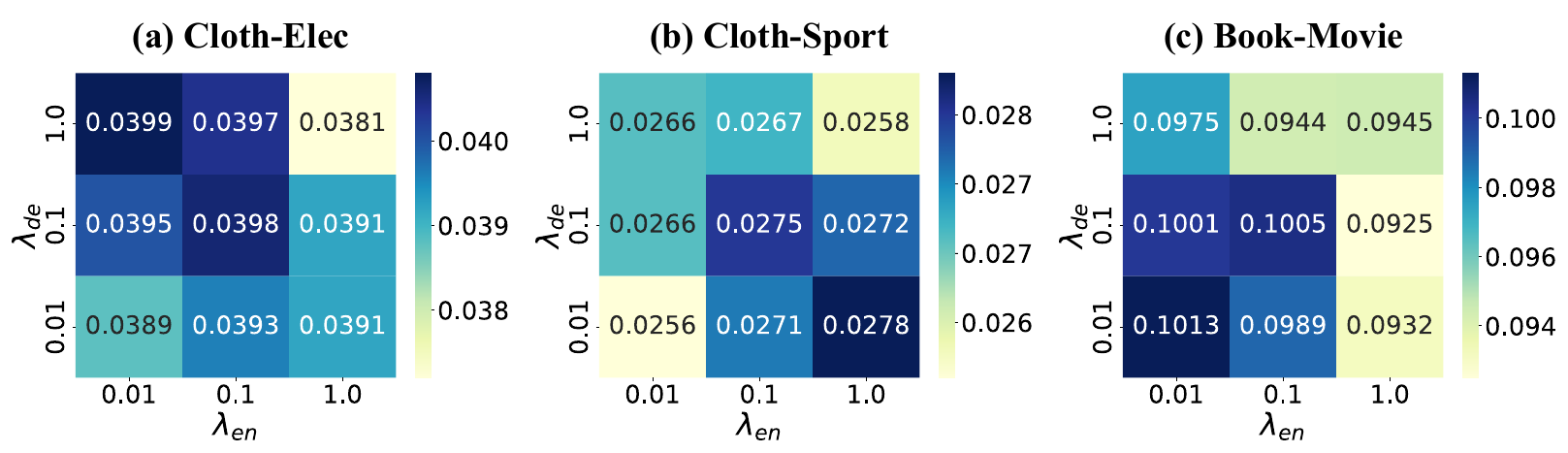}
		\Description[Recall performance under different encoder and decoder regularization strengths]
		{Three heatmaps showing Recall scores under varying combinations of $\lambda_{en}$ (encoder regularization) and $\lambda_{de}$ (decoder regularization) for different domain pairs. In (a) Cloth-Elec, the best performance is achieved at $\lambda_{en}=0.01$ and $\lambda_{de}=1.0$, indicating that strong decoder but weak encoder regularization is preferred. In (b) Cloth-Sport, optimal Recall appears at $\lambda_{en}=1.0$ and $\lambda_{de}=0.01$, favoring stronger encoder regularization. In (c) Book-Movie, the highest Recall occurs when both $\lambda_{en}$ and $\lambda_{de}$ are set to 0.01, suggesting that light regularization on both components yields better representations.}
		\caption{Performance comparison w.r.t. different values of $\lambda_{en}$ and $\lambda_{de}$ for the overall objective}
		\label{fig:paraAnaly}
	\end{figure}

	As shown in Fig.\ref{fig:paraAnaly}, we analyze Recall under varying hyperparameters: for the Cloth-Elec dataset in Fig.\ref{fig:paraAnaly}(a), the highest recall occurs at $\lambda_{en}=0.01$ and $\lambda_{de}=1.0$,  indicating that weaker encoder regularization combined with stronger decoder regularization is more conducive to achieving optimal performance. In contrast, for the Cloth-Sport dataset in Fig.\ref{fig:paraAnaly}(b), optimal Recall is observed when $\lambda_{en}=1.0$ and $\lambda_{de}=0.01$, where stronger encoder regularization and weaker decoder regularization perform best. For the Book-Movie dataset in Fig.\ref{fig:paraAnaly}(c), the highest Recall is achieved when $\lambda_{en}=0.01$ and $\lambda_{de}=0.01$, suggesting that both encoder and decoder benefit from weaker regularization, as excessive constraints may impede the learning of informative representations.
	
	While our method exhibits mild sensitivity to hyperparameters, it consistently outperforms state-of-the-art unsupervised disentangled CDR methods, even under suboptimal settings. This highlights the effectiveness of supervision signals in enhancing model stability and robustness. Nevertheless, we recognize that the dependence on hyperparameter tuning may limit the practical applicability in dynamic environments. To address this, future work will explore adaptive strategies for hyperparameter optimization, aiming to reduce manual intervention and enhance model generalizability.

	\section{Conclusion and future work}
	This paper proposes DGCDR, a cross-domain recommendation method that introduces supervision into the disentangled representation learning through an encoder-decoder framework. By leveraging GNNs to capture high-order collaborative signals, DGCDR generates enriched representations as a robust foundation for disentanglement. The anchor-based mechanism in the encoder-decoder framework effectively supervises the disentangling of intra-domain features and cross-domain alignment. These findings highlight the importance of integrating disentanglement with supervision, addressing two key challenges in existing CDR methods: the expressiveness limitations of pre-separation strategies, which disrupt the collaborative modeling of domain-specific and domain-shared features, and the difficulty of effective disentanglement in unsupervised settings, especially for cross-domain alignment. Experimental results on real-world datasets show that DGCDR outperforms existing methods in recommendation accuracy, feature disentanglement, and alignment. Future work will focus on adapting disentangled learning to evolving user behaviors and domain shifts.
	\begin{acks}
	The authors Yuhan Wang, Qing Xie, Mengzi Tang, Lin Li, and Yongjian Liu are supported partially by National Natural Science Foundation of China (Grant No. 62276196) and National Key Research and Development Program of China (2024YFF0907002).
	\end{acks}
	
	\bibliographystyle{ACM-Reference-Format}
	\bibliography{my-sample-recsys25}
\end{document}